\documentclass[preprint]{aastex}

\begin{document}

\title{Current Star Formation in Post-Starburst Galaxies?}
\author{Neal A. Miller\altaffilmark{1}} 
\affil{National Radio Astronomy Observatory\altaffilmark{2}, P.O. Box O, \\ Socorro, New Mexico  87801}
\email{nmiller@aoc.nrao.edu}

\and

\author{Frazer N. Owen}
\affil{National Radio Astronomy Observatory\altaffilmark{2}, P.O. Box O, \\ Socorro, New Mexico 87801}
\email{fowen@aoc.nrao.edu}

\altaffiltext{1}{and New Mexico State University, Department of Astronomy, Box
30001/Dept. 4500, Las Cruces, New Mexico 88003}

\altaffiltext{2}{The National Radio Astronomy Observatory is a facility of the National Science Foundation operated under cooperative agreement by Associated Universities, Inc.}

\begin{abstract}
Radio continuum observations are a probe of star formation in galaxies, and are unaffected by dust extinction. Observations of the distant rich cluster Cl 0939+4713 have detected radio galaxies classified as post-starburst (``k+a'') on the basis of their optical spectra, and presumably this situation arises from heavily dust-obscured star formation (Smail et al. 1999). We present the results of a radio continuum survey of post-starburst galaxies identified from the Las Campanas Redshift Survey by Zabludoff et al. (1996). This sample was selected using very stringent criteria, and therefore provides an estimate on the incidence of potential star formation in galaxies whose optical spectra exhibit the strongest post-starburst features. We detected two of fifteen such galaxies at radio luminosities consistent with moderate levels of star formation. This result underscores the potential importance of dust extinction when investigating star formation in galaxies.
\end{abstract}

\section{Introduction}

\citet{dres1983} spectroscopically identified an unusual population of galaxies in intermediate redshift clusters. The spectra of these galaxies exhibited both the absorption lines standard to an older stellar population as well as deep Balmer absorption indicative of a population of A stars. Since the emission lines associated with current star formation were absent from the spectra yet the spectra possessed obvious evidence of a relatively young population, the galaxies were dubbed `E+A' (i.e., an elliptical spectrum plus A stars) and surmised to be galaxies in which an episode of star formation had recently ended ($\lesssim 1$ Gyr).

Subsequent studies of these galaxies have demonstrated that most of the galaxies with E+A spectra were actually disk galaxies \citep[e.g.,][]{fran1993,couc1994,dres1994}. This led \citet{fran1993} to propose the more neutral classification of `K+A' to remove the morphological implication of E+A, which has further evolved into `k+a' in more recent studies \citep[e.g.,][]{dres1999}. It is this nomenclature that we will adopt for this Letter.

Recent observations of moderate redshift clusters have challenged the interpretation that all galaxies with k+a spectra are truly post-starburst. Deep radio images of Cl 0939+4713 detected five k+a galaxies, and HST near-infrared and optical observations of the galaxies suggest large amounts of dust \citep{smai1999}. Since radio emission is an indicator of current star formation \citep[e.g., see][]{cond1992}, the authors postulated that star formation was ongoing in some k+a's and that optical signatures of such star formation were heavily obscured by dust. The potential role of dust extinction in the creation of k+a galaxies had been noted previously. \citet{kenn1995} noted that a number of nearby galaxy mergers had spectral characteristics of k+a galaxies, but with H$\alpha$ emission. The modeling work of \citet{pogg1999} suggested that the k+a's were the product of a phase of dusty star formation also prevalent in the intermediate clusters \citep[classified as `e(a)' galaxies; see][for a description of these spectral classifications]{dres1999}, and \citet{pogg2000} further investigated this by performing spectral observations of very luminous infrared galaxies. They found evidence for selective dust extinction whose effect is dependent on stellar age \citep[also suggested by][]{calz1994}, such that the youngest stars inhabit very dusty star forming regions while older stars have had time to migrate out of such regions. \citet{bekk2001} modelled galaxy-galaxy mergers including the effects of dust extinction, confirming that such systems can produce e(a) spectra and that these galaxies evolve into galaxies with k+a spectra.

\citet{zabl1996} identified a sample of relatively nearby k+a galaxies from the Las Campanas Redshift Survey \citep[LCRS;][]{shec1996}. After making cuts by redshift and S/N, they isolated 21 k+a galaxies from an initial sample of over 11,000 galaxies. Since there is often substantial variation in the definition of the terms post-starburst and k+a, they applied a conservative definition designed to reject emission line galaxies (EW([OII]) $<$ 2.5$\mbox{\AA}$) and select only the stronger Balmer absorption systems (EW($\langle$H$\beta$ H$\gamma$ H$\delta \rangle$) $>~5.5\mbox{\AA}$). This sample was found to be predominantly field in nature, demonstrating that the k+a phenomenon is not limited to clusters. Furthermore, they noted the high incidence of merger systems in the sample and supported this origin for k+a galaxies.

The LCRS k+a sample provides an excellent case study for possible current star formation in post-starburst galaxies. Their rigorous definition based on a high average EW of three Balmer lines implies that such galaxies are among the strongest examples of post-starburst galaxies. Did the starbursts that produced the strong Balmer absorption features in these galaxies terminate entirely, or are they still forming stars at some rate? The answer to such a question is obviously important to understanding their evolutionary history, and thereby the mechanisms responsible for their creation.

We have used the Very Large Array (VLA) to collect 20cm radio continuum observations to determine whether any of the LCRS k+a galaxies show signatures of active star formation. The advantage to using radio emission as a star formation indicator is that it is unaffected by the dust obscuration that could complicate optical analysis. Consistent with the findings in higher redshift clusters, we find that at least some of these post-starburst galaxies are probably forming stars currently.

\section{Observations}

The observational goal was to probe down to low equivalent star formation rates (SFR). The relationship between SFR and radio luminosity is
\begin{equation}\label{sfr}
\mbox{SFR}~[\mbox{M}_\odot~\mbox{yr}^{-1}]~=~5.9 \times 10^{-22}~\mbox{L}_{1.4GHz}[\mbox{W}~\mbox{Hz}^{-1}]
\end{equation}
\citep{yun2001}. This conversion assumes a Salpeter initial mass function integrated over all stars ranging from 0.1 M$_\odot$ to 100 M$_\odot$, and hence represents the total SFR of a galaxy. The primary contribution to the radio luminosity is non-thermal (synchrotron), and is believed to arise from cosmic ray electrons accelerated by supernovae of massive stars. Our radio observations were designed to detect the k+a galaxies (at 3$\sigma$) if they had radio luminosities $\ge 2 \times 10^{21}$ $h_{75}^{-2}$W Hz$^{-1}$. This luminosity corresponds to a SFR of 1.2 M$_\odot$ yr$^{-1}$ if we adopt Equation \ref{sfr}.

While 20cm continuum observing with the VLA is generally straightforward, the Southern declinations of the LCRS k+a galaxies made the observations challenging. Several of the identified k+a galaxies have $\delta\sim-45^\circ$ (J2000). The VLA is located at about $34^\circ$ North latitude and has an imposed hardware limit of $10^\circ$ in elevation, which means that the southernmost k+a's are observable only for a very short period of time during transit. This problem is exacerbated in 20cm observations by the rapidly increasing system temperature of the VLA with decreasing elevation \citep[see Figure 2 of][]{perl1998}.

The experiment was designed to make use of a single track of time in each of the B and BnA configurations of the VLA. Several of the sample of 21 k+a's would have required observations on multiple days to achieve the desired sensitivity, and consequently were not observed. This reduced the sample to 15 galaxies, four of which were observed on 13 October 1999 during the BnA configuration and 11 of which were observed on 21 January 2000 during the B configuration. In each case, the VLA was set up in line mode with two IFs each consisting of seven 3.125 MHz channels for both polarizations. This setup greatly reduces bandwidth smearing and added noise from confusing sources located far from the field center. Integrations ranged from $\sim$20 minutes to $\sim$2 hours, depending on source redshift and declination. Data reduction was performed in AIPS using the standard steps.

\section{Results and Discussion}

Two of the k+a galaxies were detected at high significance, EA12 at over 14$\sigma$ and EA19 at about 6$\sigma$ (see Table \ref{tbl-1}). The positional separations between the radio detections and the optical positions listed in \citet{zabl1996} were 0.8$\arcsec$ and 2.0$\arcsec$, respectively. Based on the densities of detected radio sources and optical sources in the approximate magnitude range of the galaxies, the probabilities that such detections arise by chance superposition are under $0.1\%$. In each case, the radio source was unresolved implying the emission originates from physical scales of $\lesssim$5 $h_{75}^{-1}$kpc and $\lesssim$15 $h_{75}^{-1}$kpc, respectively\footnote{The declination of EA19 produces a very non-circular elongated beam.}.

The radio luminosities for the galaxies imply SFRs of 5.9 and 2.2 M$_\odot$ yr$^{-1}$, respectively (see Equation \ref{sfr}). While an active galactic nuclei origin for their radio emission can not be ruled out entirely, the radio luminosity function is dominated by star formation at these luminosities \citep{cond1989}. Thus, it is highly likely that these two galaxies are still in the process of forming stars and that optical evidence of such star formation is obscured. The remainder of the galaxies were not detected. While Table \ref{tbl-1} presents upper limits on their radio luminosities and corresponding SFRs, none of these galaxies represented a marginal detection at even the 2$\sigma$ level. The sensitivity of the radio observations limits the star formation in the non-detected galaxies to very low levels, with SFR $\lesssim1$ M$_\odot$ yr$^{-1}$.

It should be noted that the quoted SFRs are dependent on the adopted relationship between this quantity and radio continuum luminosity (i.e., the constant of proportionality in Equation \ref{sfr}). \citet{cond1992} assesses this relationship empirically, using the total flux density and supernova rate of the Milky Way. The supernova rate is then converted to a SFR by assuming an IMF and minimum stellar mass at which stars are massive enough to undergo supernova. \citet{yun2001} derive the relationship by assessing the SFR density in the local universe ($z \lesssim 0.15$) from the far-infrared density. They then rely on the far-infrared/radio correlation to empirically derive Equation \ref{sfr}, which predicts SFRs about a factor of two less than those derived using the relationship from \citet{cond1992}. 

This result strengthens the conclusion of \citet{smai1999} that at least some `post-starburst' galaxies are still actively forming stars. The case in which the star formation has just ceased is unlikely, as the expected lifetime of the radio emission in such a scenario is very short, a few $\times 10^7$ years. This suggests that the classification of galaxies on the basis of optical spectra collected only in the blue is thereby imprecise, as galaxies with nearly identical spectra may or may not show evidence for star formation. Indeed, \citet{kenn1995} noted this danger when assessing their merger sample as they identified galaxies with k+a spectra in the blue but strong H$\alpha$ emission. In assessing k+a galaxies it is therefore very important to acquire additional data, be it broader wavelength coverage in the optical spectrum or observations in radio continuum. This is underscored by the conservative selection used in constructing the LCRS k+a sample, as such objects were assumed to be the strongest examples of post-starburst phenomena. We do note, however, that the radio luminosities and upper limits found in this study are an order of magnitude fainter than those found in \citet{smai1999}. Thus, the associated SFRs and limits thereof for the LCRS sample are more consistent with normal to low SFRs than with large bursts of star formation.

Under the assumption that the radio emission is due to star formation, we may also determine a gross estimate of the extinction required to reduce the equivalent width of the [OII] line to less than 2.5. \citet{kenn1998} presents a relationship between [OII] emission and star formation of  
\begin{equation}\label{oiisfr}
\mbox{SFR}~[\mbox{M}_\odot~\mbox{yr}^{-1}]~=~(1.4 \pm 0.4) \times 10^{-41}~\mbox{L}_{\mbox{[OII]}}[\mbox{erg}~\mbox{s}^{-1}].
\end{equation}
This equation may be used to predict the expected [OII] line luminosities in the absence of extinction if we adopt the SFRs implied by the radio emission. To relate this to an equivalent width, we use as a template the continuum flux of an e(a) galaxy of comparable magnitude and distance \citep[from][]{mill2001}. In the case of EA12 an extinction at H$\alpha$ of nearly 5 magnitudes is required to reduce the equivalent width to under 2.5$\mbox{\AA}$. For EA19, the corresponding calculation yields an extinction of about 3.5 magnitudes. \citet{kenn1998} notes that SFRs calculated using [OII] are imprecise as there is a wide variation in observed [OII]/H$\alpha$ ratio in galaxies, up to an order of magnitude. Should the radio SFR and [OII] SFR differ by an order of magnitude, the extinctions would still be somewhat higher than those observed in normal galaxies, at just over 2 magnitudes for EA12.

It is therefore misleading to think of the k+a class as one in which star formation has completely halted. Within this optically-defined category there may exist a broad range of SFRs, possibly including starbursts that reside in regions of very heavy dust extinction. The k+a galaxies with radio emission may then represent a bridge between the dusty starbursts of e(a) spectral classification and the true post-starburst galaxies. If this is the case, their locations within clusters would present a powerful tool in understanding the environmental conditions which create both the dusty starbursts and post-starburst galaxies.

\acknowledgments
The authors thank Barry Clark for an excellent job in scheduling these observations. The time slots fit the challenging observational constraints perfectly. We also benefitted from conversations with Ann Zabludoff, Jacqueline van Gorkum, and TzuChing Chang. N.A.M. would like to thank the NRAO predoctoral program for support of this research.

\clearpage

\begin{deluxetable}{l c c c c c c}
\tablecaption{Source Properties \label{tbl-1}} 
\tablewidth{0pt}
\tablehead{
\colhead{} & \colhead{} & \colhead{} & \colhead{rms} &
\colhead{flux} & \colhead{$\log($P$_{1.4GHz})$} & SFR \\
\colhead{Source} & \colhead{z} & \colhead{$M_R$} & 
\colhead{[$\mu$Jy beam$^{-1}$]} & \colhead{[mJy]} & 
\colhead{[W Hz$^{-1}$]} & \colhead{[M$_\odot$ yr$^{-1}$]}
}
\startdata
EA1 & 0.0747 & -19.9 & \phn82.6 & $\leq$0.248 & $\leq$21.4 & $\leq$1.4 \\
EA3 & 0.0811 & -22.1 & \phn64.2 & $\leq$0.193 & $\leq$21.3 & $\leq$1.3 \\
EA6 & 0.0885 & -20.5 & \phn54.5 & $\leq$0.163 & $\leq$21.3 & $\leq$1.3 \\
EA8 & 0.1122 & -20.7 & \phn34.3 & $\leq$0.103 & $\leq$21.3 & $\leq$1.3 \\
EA9 & 0.0651 & -19.2 & \phn74.5 & $\leq$0.223 & $\leq$21.2 & $\leq$1.0 \\
EA12 & 0.0971 & -20.7 & \phn44.5 & \phm{$\leq$}0.614 & \phm{$\leq$}22.0 & \phm{$\leq$}5.9 \\
EA13 & 0.0957 & -21.6 & \phn57.5 & $\leq$0.172 & $\leq$21.4 & $\leq$1.6 \\
EA14 & 0.0704 & -20.8 & \phn74.6 & $\leq$0.224 & $\leq$21.3 & $\leq$1.2 \\
EA15 & 0.1138 & -20.6 & \phn32.9 & $\leq$0.099 & $\leq$21.3 & $\leq$1.3 \\
EA16 & 0.0764 & -20.3 & \phn65.0 & $\leq$0.195 & $\leq$21.3 & $\leq$1.2 \\
EA17 & 0.0609 & -19.5 & \phn96.8 & $\leq$0.290 & $\leq$21.3 & $\leq$1.1 \\
EA18 & 0.0598 & -20.4 & \phn84.9 & $\leq$0.255 & $\leq$21.2 & $\leq$1.0 \\
EA19 & 0.0640 & -20.2 & \phn81.5 & \phm{$\leq$}0.501 & \phm{$\leq$}21.6 & \phm{$\leq$}2.2 \\
EA20 & 0.0632 & -20.6 & 116.\phn & $\leq$0.348 & $\leq$21.4 & $\leq$1.5 \\
EA21 & 0.0994 & -20.6 & \phn41.7 & $\leq$0.125 & $\leq$21.3 & $\leq$1.3 \\
\enddata

\tablecomments{Source names, redshifts, and apparent magnitudes from \citet{zabl1996}. All absolute quantities have been calculated assuming H$_o$ = 75 km s$^{-1}$ Mpc$^{-1}$, and SFRs using Equation \ref{sfr}.}
\end{deluxetable}

\end{document}